\documentclass[aps,prmaterials,twocolumn,superscriptaddress,reprint,citeautoscript]{revtex4-2}
\usepackage{etoolbox}
\usepackage[version=4]{mhchem}
\usepackage{graphicx}
\usepackage{dcolumn}
\usepackage{bm}

\usepackage{url}

\usepackage[usenames,dvipsnames]{xcolor}

\begin{document}

\title{
\textsc{RevelsMD}: Reduced Variance Estimators of the Local Structure in Molecular Dynamics} 

\author{Samuel W. Coles }
\affiliation{Department of Chemistry, University of Bath, Claverton Down BA2 7AY, United Kingdom}
\affiliation{The Faraday Institution, Quad One, Harwell Science and Innovation Campus, Didcot OX11 0RA, United Kingdom}

\author{Benjamin J. Morgan}
\affiliation{Department of Chemistry, University of Bath, Claverton Down BA2 7AY, United Kingdom}
\affiliation{The Faraday Institution, Quad One, Harwell Science and Innovation Campus, Didcot OX11 0RA, United Kingdom}

\author{Benjamin Rotenberg}
\affiliation{Sorbonne Universit\'{e}, CNRS, Physico-chimie des \'{E}lectrolytes et Nanosyst\`{e}mes Interfaciaux, PHENIX, F-75005 Paris, France}
\affiliation{R\'{e}seau sur le Stockage \'{E}lectrochimique de l'\'{E}nergie (RS2E), FR CNRS 3459, 80039 Amiens Cedex,
France}

\date{\today}%



\begin{abstract}
\textsc{RevelsMD} is a new open source Python library, which uses reduced variance force sampling based estimators to calculate 3D particle densities and radial distribution functions from molecular dynamics simulations. This short note describes the scientific background of the code, its utility and how it fits within the current zeitgeist in computational chemistry and materials science.
\end{abstract}
\maketitle

\section*{Summary}

Two common equilibrium structural descriptors calculated from molecular dynamics (MD) simulations are time-average particle densities and radial distribution functions (RDFs).
These descriptors are usually calculated by taking histograms of positions~\cite{Rotenberg_2020}, for particle densities, or particle--particle separations, for RDFs. Over the past ten years, force sampling, where both atom positions and the instantaneous forces acting on these atoms are used to estimate equilibrium structural descriptors, has emerged as an alternative to simple histogram methods~\cite{Rotenberg_2020,Borgis_2013,de_las_Heras_2018,Schultz_2019,Schultz_2016,Coles_2019,Coles_2021,Renner_2023}.
In particular, force-sampling methods exhibit greatly reduced variance (reduced sampling noise) when compared to a direct histogram calculated from the same microscopic configurations.
In cases where estimated 3D particle densities or RDFs are used for simple qualitative or pictoral analysis, the benefit of using force-sampling methods is mainly aesthetic, as these give smoother output data for a given input dataset.
When these equilibrium structural descriptors, however, are used as inputs for further quantitative analysis, such as calculation of excess entropy from RDFs~\cite{Ghaffarizadeh_2023}, using force-sampling methods for the initial analysis results in reduced statistical uncertainty in any subsequently derived quantities of interest.
In addition, the high energy and carbon cost of production MD simulations makes it incumbent on the modelling community to ``do more with less'' by developing and using post-processing methods that yield results in a statistically efficient fashion, and hence require smaller MD datasets to produce scientifically useful results~\cite{Frenkel_Daan2023-07-13,McCluskeyEtAl_arXiv2023}.

\textsc{RevelsMD} (Reduced Variance Estimators of the Local Structure in Molecular Dynamics)\cite{RevelsMD} is a Python implementation of the reduced-variance force-sampling methods first descibed by Borgis \textit{et al.}~\cite{Borgis_2013} and later expanded in Refs.\,\citenum{Coles_2019,Coles_2021}. Using these techniques \textsc{RevelsMD} allows the user to calculate reduced variance RDFs and 3D densities from trajectories from a variety of MD codes.

While the mathematical intricacies of force-based reduced variance estimators vary, at their core they have a common idea: calculating a property of interest by ``integrating its gradient''\cite{Rotenberg_2020}.
More precisely each method calculates the quantity of interest using the forces acting on the atoms in addition to their positions.
This avoids the noise associated with the ideal contribution to density while focusing on the the non-trivial changes in density caused by interactions. 

As an example of this approach, we consider the force-sampling method for the calculation of 3D densities first described in Ref.\,\citenum{Borgis_2013}.
The 3D number density at position $\mathbf{r}$, $\rho(\mathbf{r})$, is defined as:
\begin{equation}
\rho(\mathbf{r})=\left\langle\sum_{i=1}^N \delta\left(\mathbf{r}_i-\mathbf{r}\right)\right\rangle \;,  
\end{equation}
where the sum runs over atoms, the angular brackets denote an ensemble average, and $\delta$ is the Dirac delta function. The gradient of this density is proportional to the force density 
\begin{equation}
F(\mathbf{r})=\left\langle\sum_{i=1}^N \delta\left(\mathbf{r}_i-\mathbf{r}\right) \mathbf{f}_i\right\rangle \;, 
\end{equation}
where $\mathbf{f}_i$ is the force acting on atom $i$, with a proportionality constant $\beta=1/k_BT$, where $k_B$ is the Boltzmann constant and $T$ the temperature. This leads to the alternative estimator of the density:
\begin{equation}
    \rho(\mathbf{k})=-\frac{i \beta}{k^2} \mathbf{k} \cdot \mathbf{F}(\mathbf{k}) -\rho_{0} \; ,
\end{equation}
where $F(\mathbf{k})$ is the force density in $k$-space and can be obtained by performing a Fast Fourier Transform on the force density, and the real space number density obtained using an inverse transform. 

In addition to calculating force based RDFs and 3D densities \textsc{revelsMD} also uses these estimators to calculate optimal linear combinations of two estimators using a control variates approach\cite{HAMMER_2008}. Here a position dependent linear combination ($E_\lambda(x)$) is obtained for a descriptor from two different estimators, $E_{0}(x)$ and $E_{1}(x)$,
\begin{eqnarray}
    E_\lambda(x)&=&(1-\lambda(x)) E_0(x)+\lambda(x) E_1(x) \nonumber  \\
    &=& E_0(x)+\lambda(x)\Delta(x) \, ,
\end{eqnarray}
where $\Delta(x)= E_1(x)- E_0(x)$. While $\lambda(x)$ can have any value, using a control variates approach we calculate the value which minimises the variance of $E_\lambda(x)$ at each point as~\cite{Coles_2021},
\begin{equation}
    \lambda(x)=-\frac{\textrm{Cov}\left(E_0(x), \Delta(x)\right)}{\textrm{Var}(\Delta(x))}
\end{equation}
where $\textrm{Var}$ and $\textrm{Cov}$ are the variance and covariance operators, respectively.. \textsc{RevelsMD} applies this method to two different force-based estimators in its RDF module as in Ref.~\citenum{Coles_2021}, while the 3D density control variates implementation combines a force-based and a conventional histogram-of-positions estimator.

\section*{Statement Of Need}

\textsc{RevelsMD} is the first publicly available code that uses force-sampling methods to calculate 3D particle densities and radial distribution functions. These two equilibrium structural descriptors play a fundamental role throughout computational chemistry, in fields that include molecular solvation~\cite{Shimizu_2015,2013}, biophysics~\cite{Lemkul_2019,Coles_2019} and materials science~\cite{Cherry_1995,Mercadier_2023}.
The last year has been a landmark year for this family of methods, with two key developments taking place that indicate that they are ready for broader use. First, these methods have been applied for the first time to a real chemical system, where they were used to analyse ion-conduction channels in the fluoride solid electrolyte cubic \ce{BaSnF4}~\cite{Mercadier_2023}. Second, the previously mentioned imperative to ``do more with less" was highlighted in the third edition of ``Understanding Molecular Simulation'' ~\cite{Frenkel_Daan2023-07-13}, with this family of methods noted as exemplifying this principle. 

3D particle densities and radial distribution functions are both used to understand equilibrium structure and to calculate statistical mechanical quantities.
In both cases, using an estimator with reduced variance means a smaller number of simulation configurations are needed to reach a given precision in the resulting descriptor.
This allows these structural properties of interest to be obtained at reduced overall computational cost, and, in some cases, allows scientifically meaningful results to be obtained in cases where the cost of simulation is too high to use computational brute force to converge conventional histogram-based estimators, e.g., where molecular dynamics is run using electronic structure methods to calculate forces (so-called ab initio molecular dynamics, AIMD). 

\textsc{RevelsMD}, can calculate reduced-variance estimators directly from the output of the \textsc{lammps}, and \textsc{vasp} codes, and also has generic \textsc{MDanalysis} and \textsc{numpy} based interfaces for more general compatibility with generic simulation workflows.   
\newline

\section*{Acknowledgements}

We would like to thank Daniel Borgis, Rodolphe Vuilleumier, Etienne Mangaud and Daan Frenkel for their contributions over the past decade, which were essential to the development of \textsc{RevelsMD}. This project received funding from the European Research Council under the European Union's Horizon 2020 research and innovation program (Grant Agreement No. 863473), and from the Faraday Institution CATMAT project (EP/S003053/1, FIRG016).

\bibliography{JOSS}

\end{document}